\documentclass[10pt,twocolumn]{article}

\usepackage[utf8]{inputenc}
\usepackage{amsmath}
\usepackage{amssymb}
\usepackage{amsthm}
\usepackage{titlesec}
\usepackage{cite}
\usepackage{color}
\usepackage{booktabs}
\usepackage{graphicx}
\usepackage[colorlinks=false]{hyperref}
\usepackage[format=plain,labelfont=it]{caption}
\usepackage[left=1.5cm,right=1.5cm,top=2cm,bottom=2cm]{geometry}

\titleformat{\section}{\centering\normalfont\scshape}{\Roman{section}.}{5pt}{}
\titleformat{\subsection}{\normalfont\it}{\Alph{subsection}.}{5pt}{}
\titleformat{\subsubsection}{\normalfont\it}{\hspace{4mm}\arabic{subsubsection})}{5pt}{}

\newcommand\infoFootnote[1]{%
  \begingroup
  \renewcommand\thefootnote{}\footnote{#1}%
  \addtocounter{footnote}{-1}%
  \endgroup}

\newtheorem{thm}{Theorem}
\newtheorem{cor}[thm]{Corollary}
\newtheorem{lem}[thm]{Lemma}

\newtheorem{prob}{Problem}

\newcommand{\R}{\mathbb{R}} 
\newcommand{\N}{\mathbb{N}} 

\newcommand{\Fc}{\mathcal{F}}
\newcommand{\Rc}{\mathcal{R}}
\newcommand{\Tc}{\mathcal{T}} 
\newcommand{\Uc}{\mathcal{U}} 
\newcommand{\Xc}{\mathcal{X}} 

\newcommand{\Ab}{\boldsymbol{A}}
\newcommand{\Bb}{\boldsymbol{B}}

\newcommand{\Kb}{\boldsymbol{K}}
\newcommand{\Lb}{\boldsymbol{L}}

\newcommand{\Pb}{\boldsymbol{P}}
\newcommand{\Qb}{\boldsymbol{Q}}
\newcommand{\Rb}{\boldsymbol{R}}
\newcommand{\Sb}{\boldsymbol{S}}

\newcommand{\Wb}{\boldsymbol{W}}

\newcommand{\ab}{\boldsymbol{a}}
\newcommand{\bb}{\boldsymbol{b}}

\newcommand{\fb}{\boldsymbol{f}}
\newcommand{\gb}{\boldsymbol{g}}
\newcommand{\hb}{\boldsymbol{h}}

\newcommand{\lb}{\boldsymbol{l}}

\newcommand{\ub}{\boldsymbol{u}}

\newcommand{\xb}{\boldsymbol{x}}
\newcommand{\yb}{\boldsymbol{y}}
\newcommand{\zb}{\boldsymbol{z}}

\newcommand{\xib}{\boldsymbol{\xi}}
\newcommand{\pib}{\boldsymbol{\pi}}
\newcommand{\Phib}{\boldsymbol{\Phi}}
\newcommand{\zerob}{\boldsymbol{0}}

\newcommand{\blind}[1]{\textcolor{white}{#1}}
\newcommand{\interior}{\mathrm{int}}

\DeclareMathOperator*{\argmin}{arg\,min}

\title{\vspace{-2mm}\bf Tailored neural networks for learning optimal value functions in MPC}
\author{Dieter Teichrib and Moritz Schulze Darup\vspace{2mm}}
\date{}

\begin{document}

\maketitle

\textbf{\textit{Abstract}.} {\bf Learning-based predictive control is a promising alternative to optimization-based MPC. However, efficiently learning the optimal control policy, the optimal value function, or the Q-function requires suitable function approximators. Often, artificial neural networks (ANN) are considered but choosing a suitable topology is also non-trivial.  
Against this background, it has recently been shown that tailored ANN allow, in principle, to exactly describe the optimal control policy in linear MPC by exploiting its piecewise affine structure.
In this paper, we provide a similar result for representing the optimal value function and the Q-function that are both known to be piecewise quadratic for linear MPC.}
% leave no space here
\infoFootnote{D. Teichrib and M. Schulze Darup are with the \href{https://rcs.mb.tu-dortmund.de/}{Control and~Cyber-physical Systems Group}, Faculty of Mechanical Engineering, TU Dortmund University, Germany. E-mails:  \href{mailto:moritz.schulzedarup@tu-dortmund.de}{\{dieter.teichrib, moritz.schulzedarup\}@tu-dortmund.de}. \vspace{0.5mm}}
\infoFootnote{\hspace{-1.5mm}$^\ast$This paper is a \textbf{preprint} of a contribution to the 60th IEEE Conference on Decision and Control 2021.}

\section{Introduction}

Learning-based predictive control becomes more and more prominent in research and practice. Various approaches have been presented that underline the capabilities of (deep) learning in the context of control (see, e.g., \cite{Bradtke1994,Lewis2009,Domahidi2011,Zhong2013,Chen2018,Bhardwaj2020,Karg2020}). Most of these approaches either aim for approximations of the optimal control law in the policy space  (such as \cite{Domahidi2011,Karg2020}) or the optimal value function (or Q-function) in the value space (such as \cite{Zhong2013,Bhardwaj2020}). In both cases, suitable parametrizations for the functions to be learned have to be chosen. Common candidates are, for example, artificial neural networks (ANN) or Bayesian networks.

Choosing the type of parametrization and underlying specifications (such as layer width and depth or activation functions in ANN) is often done in a trial-and-error fashion. However, some setups support more ``educated guesses''. In fact, it has recently been shown that  certain types of ANN allow, in principle, to exactly describe model predictive control (MPC) laws for linear systems \cite{Karg2020,SchulzeDarup_2020_ECC_ANN}. The key observation here is that both the MPC law and certain ANN reflect piecewise affine (PWA) functions on polyhedral partitions. More specifically, feed-forward ANN with rectified linear units (ReLU) or max-out activations \cite{Goodfellow2013} offers this feature \cite{Montufar2014}.   

Now, the optimal value function (OVF) in linear MPC is well-known to be piecewise quadratic (PWQ) \cite{Bemporad2002}. The main contribution of this paper is to show that suitable ANN can also be found for such functions. Interestingly, these ANN will also build on ReLU activations. However, successful realizations require to extend the input layer by quadratic terms of the system states (such as $\xb_1^2$, $\xb_2^2$, and $\xb_1\xb_2$). Extended inputs of this form  have also been considered elsewhere. For instance, \cite{Cheung1991} and \cite{Fan2019} exploit their ability to create non-linear decision boundaries. Clearly, this feature is not helpful here since we are dealing with polyhedral partitions. In contrast, we will use the quadratic terms to eliminate the quadratic expressions of the OVF by suitably choosing the affine preactivations. For the remaining PWA structure, we can exploit the known results mentioned above. In combination, we obtain an approach for exactly describing the OVF via ANN with ReLU activations. In addition, we present a similar approach for representing Q-functions.

The paper is organized as follows. 
In Section~\ref{sec:background}, we collect preliminaries on ANN, MPC, and learning-based predictive control. Further, we briefly summarize the known representation of PWA functions using ANN. We present our main result, i.e., tailored ANN for exactly describing PWQ functions, in Section~\ref{sec:topologies}. More specifically, we construct ANN that exactly describe the OVF in linear MPC for the special case of one-dimensional states (i.e., $n=1$). Moreover, we show that ANN for describing the Q-function can be derived whenever a construction for the OVF is known  (for any $n$).
Finally, we illustrate our results with various numerical examples in Section~\ref{sec:example} and state conclusions in Section~\ref{sec:conclusion}.

\section{Preliminaries and background}
\label{sec:background}

\subsection{Neural networks with rectifier activation}
\label{subsec:neuralNetworks}

In general, a feed-forward ANN with $l\in\N$ hidden layers and $w_i$ neurons in layer $i$ can be written as a composition of the form
\begin{equation}\label{eq:ANN}
\Phib(\xib)=\fb^{(l+1)}\circ \gb^{(l)}\circ \fb^{(l)}\circ \dots \circ \gb^{(1)}\circ \fb^{(1)}(\xib),
\end{equation}
where the functions $\fb^{(i)}: \R^{w_{i-1}} \rightarrow \R^{w_i}$ refer to preactivations (for $i\leq l$) and a postactivation (for $i=l+1$) and where $\gb^{(i)}: \R^{w_i} \rightarrow \R^{w_i}$ are activation functions.
The pre- and postactivations are typically affine, i.e.,
\begin{equation}
\nonumber
\fb^{(i)}(\yb^{(i-1)})=\Wb^{(i)}\yb^{(i-1)}+\ab^{(i)},
\end{equation}
where $\Wb^{(i)}\in\R^{w_i\times w_{i-1}}$ is a weighting matrix, $\ab^{(i)} \in \R^{w_i}$ is a bias vector, and 
 $\yb^{(i-1)}$ denotes the output of the previous layer with $\yb^{(0)}:=\xib$. 
Now, while various activation functions are established, we here focus on activations via ReLU that are characterized by
\begin{equation}
\nonumber
\gb^{(i)}(\zb^{(i)})=\max\left\{\zerob,\zb^{(i)}\right\}:=\begin{pmatrix}
\max\big\{0,\zb_1^{(i)}\big\} \\
\vdots \\
\max\big\{0,\zb_{w_i}^{(i)}\big\}
\end{pmatrix}.
\end{equation}
We will refer to the resulting networks as ReLU-ANN.

\subsection{Model predictive control and structural insights}

Classical MPC builds on solving an optimal control problem (OCP) of the form
\begin{align}
\label{eq:OCP}
V_N(\xb) := \!\!\!\!\min_{\substack{\hat{\xb}(0),...,\hat{\xb}(N)\\ \hat{\ub}(0),...,\hat{\ub}(N-1)}} 
\!\!\!\!\!\!\!\!\!\!\!\!\!\!\!\!\!\!\!\! & 
\,\,\,\,\,\,\,\,\,\,\,\,\,\,\,\,\,\,\,\,\,
\varphi( \hat{\xb}(N)) + \! \sum_{\kappa=0}^{N-1} \ell(\hat{\xb}(\kappa),\hat{\ub}(\kappa))   \\
\nonumber
\text{s.t.} \quad \quad  \hat{\xb}(0)&=\xb, \\
\nonumber
 \hat{\xb}(\kappa+1)&=\Ab\,\hat{\xb}(\kappa) + \Bb \hat{\ub}(\kappa), \quad\!\!\forall \kappa \in \{0,...,N-1\}, \\
 \nonumber
\left(\hat{\xb}(\kappa),\hat{\ub}(\kappa)\right) & \in \Xc \times \Uc, \quad\hspace{14.9mm}\forall \kappa \in \{0,...,N-1\}, \\
 \nonumber
 \hat{\xb}(N) & \in \Tc\\[-6.5mm]
 \nonumber
\end{align}
in every time step $k\in\N$ for the current state ${\xb=\xb(k)}$. 
Here, $N\in\N$ refers to the prediction horizon and 
\begin{equation}
\varphi(\xb):= \xb^\top \Pb \xb \quad \text{and} \quad \ell(\xb,\ub):=\xb^\top \Qb \xb + \ub^\top \Rb \ub
\end{equation}
denote the terminal and stage cost, respectively,  where the weighting matrices  $\Pb$, $\Qb$, and $\Rb$ are positive (semi\nobreakdash-) definite. The dynamics of the linear prediction model are described by $\Ab\in \R^{n \times n}$ and ${\Bb \in \R^{n\times m}}$. State and input constraints can be incorporated via the polyhedral sets $\Xc$ and $\Uc$. Finally, the terminal set $\Tc$ allows to enforce closed-loop stability 
(see \cite{Mayne2000} for details).
The resulting control policy $\pib:\Fc_N \rightarrow \Uc$ is defined~as
\begin{equation}
\label{eq:gMPC}
\pib(\xb):=\hat{\ub}^\ast(0),
\end{equation}
where $\Fc_N$ denotes the feasible set of~\eqref{eq:OCP} and where $\hat{\ub}^\ast(0)$ refers to the first element of the optimal input sequence.

Now, it is well known that $\pib(\xb)$ is a (continuous) PWA function of the form
\begin{equation}\label{eq:uPWA}
\pib(\xb) = \left\{
\begin{array}{cc}
 \Kb^{(1)} \xb + \bb^{(1)} & \text{if}\,\,\,\xb\in\Rc^{(1)}, \\
\vdots & \vdots \\
 \Kb^{(s)} \xb+ \bb^{(s)} & \text{if}\,\,\,\xb\in\Rc^{(s)}, \\
\end{array}
\right. 
\end{equation} 
where the regions $\Rc^{(i)}$ represent polyhedral sets with pairwise disjoint interiors \cite[Thm.~4]{Bemporad2002}. A related observation is that the optimal value function (OVF) is a (continuous and convex) piecewise quadratic (PWQ) function of the form
\begin{equation}\label{eq:VPWQ}
V_N(\xb) = \left\{
\begin{array}{cc}
\!\!\xb^\top\Sb^{(1)}\xb +\xb^\top \lb^{(1)} +c^{(1)} & \text{if}\,\,\,\xb\in\Rc^{(1)}, \\
\vdots &  \vdots \\
\!\!\xb^\top\Sb^{(s)}\xb +\xb^\top \lb^{(s)} +c^{(s)} & \text{if}\,\,\,\xb\in\Rc^{(s)}. \\
\end{array}
\right.\!
\end{equation}

\subsection{Learning-based predictive control}
\label{subsection:learning_for_MPC}

In principle, the structure \eqref{eq:uPWA} is quite beneficial for efficiently implementing MPC. In fact, for moderate system dimensions and prediction horizons, \eqref{eq:uPWA} can be computed explicitly and stored on the controller device. 
However, for larger dimensions or horizons, this procedure is intractable. If, in addition, an online solution of the OCP is computationally too demanding, machine learning (ML) is a promising alternative. Most intuitively, ML can be used to approximate \eqref{eq:uPWA}, which refers to an approximation in the policy space. 
This approach in combination with ANN of the form \eqref{eq:ANN} has, e.g., been considered in \cite{Chen2018,Karg2020}. Alternatively, one can also aim for approximating \eqref{eq:VPWQ}, i.e., an approximation in the value space \cite{Zhong2013}, \cite{Bertsekas2019book}. In fact, an (almost) optimal control input can also be inferred from (an approximation of) the OVF \cite{Lewis2009}. 
Remarkably, approximating \eqref{eq:uPWA} or \eqref{eq:VPWQ} is typically realized offline using supervised learning. Hence, it usually involves the solution of \eqref{eq:OCP} for sampled states $\xb$ and thus requires the knowledge of a model in terms of $\Ab$ and $\Bb$. Nevertheless, learning-based predictive control can also be realized without a model. To this end, one aims for approximating the so-called Q-function (see, e.g., \cite[P. 13]{Bertsekas2019book}) 
\begin{equation}\label{eq:Q-function}
Q_N(\xb,\ub):=\ell(\xb,\ub) + V_{N-1}(\Ab \xb + \Bb \ub) 
\end{equation}
that reflects the costs for a first step with an undetermined~$\ub$ followed by an optimal control sequence of length $N-1$. We note, in this context, that the successor state ${\xb^+:=\Ab \xb + \Bb \ub}$ can either be computed (using a model) or measured (without a model). Based on the Q-function, one can then easily derive an optimal input by evaluating $\argmin_{\ub} Q_N(\xb,\ub)$. Approximating~\eqref{eq:Q-function} in the context of MPC has, e.g., been considered in \cite{Bhardwaj2020}.

\subsection{Tailored neural networks for representing PWA functions}
\label{subsection:PWA_ANN}

In this paper, we aim for tailored ANN that are capable of exactly describing the PWQ OVF \eqref{eq:VPWQ} or the Q-function~\eqref{eq:Q-function}. This goal is motivated by the observation that ReLU-ANN can, in principle, exactly describe any (continuous) PWA function of the form~\eqref{eq:uPWA} (see, e.g., \cite{Arora2018}, \cite{Hanin2017}).
However, the corresponding network topologies can quickly become intractable. For instance, the approach in \cite[Thm.~2]{Hanin2017} requires at least $s$ hidden layers, which typically results in an ANN with unfavorable depth. Useful topologies can, however, be found for special cases. In fact, for $n=1$ (and arbitrary $m$), \eqref{eq:uPWA} can exactly be described by a ReLU-ANN with one hidden layer of width $w_1=s$, i.e., by a function of the form
\begin{equation}
\label{eq:ReLUoneLayer}
   \Phib(\xib)=\Wb^{(2)} \max\left\{ \zerob, \Wb^{(1)}\xib + \ab^{(1)} \right\} + \ab^{(2)} 
\end{equation}
with $\Wb^{(1)},\ab^{(1)}\in \R^s$, $\Wb^{(2)} \in \R^{m\times s}$, and $\ab^{(2)}\in \R^m$. 
Suitable weights and biases are given in \cite{SchulzeDarup_2020_ECC_ANN}. Their construction builds on the trivial observation that, for one-dimensional $\xb$, the regions $\Rc^{(i)}$ reflect proper intervals of the form
\begin{equation}
\label{eq:intervalRegions}
\Rc^{(i)}=[\underline{x}^{(i)},\overline{x}^{(i)}] \subset \R
\end{equation}
 that can be sorted such that $\overline{x}^{(i)}=\underline{x}^{(i+1)}$ for every $i\in \{1,\dots,s-1\}$. Using these conditions, equality of \eqref{eq:uPWA} and \eqref{eq:ReLUoneLayer} holds for the parametrization
\begin{align}
\label{eq:weightsBiasesPWA}
 \Wb^{(1)}\!&:=\!\begin{pmatrix}
-1 \\ 
\blind{+}1 \\ 
\blind{+}\vdots \\ 
\blind{+}1 \\ 
\end{pmatrix},
\qquad 
\ab^{(1)}:=\begin{pmatrix}
\blind{+}\overline{x}^{(1)} \\ 
-\overline{x}^{(1)} \\ 
\vdots \\ 
-\overline{x}^{(s-1)} 
\end{pmatrix}, \\
\nonumber
\Wb^{(2)}\!&:=\!\begin{pmatrix}-\Kb^{(1)} & \!\Kb^{(2)} & \Kb^{(3)}\!\!-\!\Kb^{(2)} & \!\!\!\dots\!\!\!\! &  \Kb^{(s)}\!\!-\!\Kb^{(s-1)} \end{pmatrix}\!,\\
\nonumber
\ab^{(2)}\!&:=\Kb^{(1)}\overline{x}^{(1)} + \bb^{(1)}
\end{align}
and the choice $\xib:=\xb$ \cite[Thm.~1]{SchulzeDarup_2020_ECC_ANN}. The paper now intends to derive analogue parametrizations for describing~\eqref{eq:VPWQ} and~\eqref{eq:Q-function}.

\section{\!\!Tailored neural networks for PWQ functions}
\label{sec:topologies}

\subsection{Problem specification}

As mentioned in the introduction, our goal is to design ReLU-ANN that allow to eliminate the quadratic terms in \eqref{eq:VPWQ} or~\eqref{eq:Q-function}. In fact, the remaining functions will then be PWA and, hence, we can reuse the results from Section~\ref{subsection:PWA_ANN} for describing the remainders using further ReLU-ANN.
A formalization of this concept leads to the two following problem statements.

\begin{prob}%[Tailored ANN for OVF]
\label{prob:ANNforOVF}
Identify an ANN of the form~\eqref{eq:ReLUoneLayer} and a mapping $\hb_v: \R^n \rightarrow \R^{w_0}$ such that 
\begin{equation}\label{eq:deltaV}
\Delta V_N(\xb):=V_N(\xb)-\Phib(\hb_v(\xb))
\end{equation}
is PWA in $\xb$.
\end{prob}

\begin{prob}%[Tailored ANN for $Q$-func.]
\label{prob:ANNforQfun}
Identify an ANN of the form~\eqref{eq:ReLUoneLayer} and a mapping $\hb_q: \R^n \times \R^m \rightarrow \R^{w_0}$ such that 
\begin{equation}
\label{eq:deltaQ}
  \Delta Q_N(\xb,\ub):=Q_N(\xb,\ub)-\Phib(\hb_q(\xb,\ub))  
\end{equation}
is PWA in $\xb$ and $\ub$.
\end{prob}

Here, it is import to note that choosing suitable network inputs via $\hb_v$ and $\hb_q$ is part of the problems. In fact, the naive choices $\xib:=\xb$ as in Section~\ref{subsection:PWA_ANN} or $\xib^\top:=(\,
\xb^\top \,\,\, \ub^\top \,)^\top$ will, in general, not solve the problems.
We further note that both $V_N$ and $Q_N$ are scalar-valued functions. Hence, also the desired $\Phib$ is scalar-valued in both cases and the corresponding parameters have the dimensions $\Wb^{(1)}\! \in \R^{w_1 \times w_0}$, $\ab^{(1)}\!\in \R^{w_1}$, $\Wb^{(2)}\! \in \R^{1\times w_1}$, and $\ab^{(2)}\!\in \R$. 
We still use bold characters for $\Phib$ and $\ab^{(2)}$ for consistency.

In the following, we present (partial) solutions to Problems~\ref{prob:ANNforOVF} and~\ref{prob:ANNforQfun}. More precisely, we solve Problem~\ref{prob:ANNforOVF} for the special case $n=1$ that has also been addressed in \cite{SchulzeDarup_2020_ECC_ANN} for the PWA analogue. Further, we show that a solution to Problem~\ref{prob:ANNforQfun} can be derived from a given solution to Problem~\ref{prob:ANNforOVF} for arbitrary $n\geq 1$.

\subsection{Solution to Problem 1 for $n=1$}

The following theorem provides a solution to Problem~\ref{prob:ANNforOVF} for one-dimensional system states. The underlying construction is based on the observation that ReLU can be designed to be active only inside one of the regions~\eqref{eq:intervalRegions}.

\begin{thm}
\label{thm:sol1}
Let $n=1$ and assume that the regions $\Rc^{(i)}$ are as in~\eqref{eq:intervalRegions} (with $\underline{x}^{(i)}<\overline{x}^{(i)}=\underline{x}^{(i+1)}$) and bounded. Then, a solution to Problem~\ref{prob:ANNforOVF} is given in terms of
\begin{align*}
    \Wb^{(1)}&:=\begin{pmatrix} 
\underline{x}^{(1)}+\overline{x}^{(1)} & -1 \\ 
\vdots & \vdots \\ 
\underline{x}^{(s)}+\overline{x}^{(s)} & -1 \\ 
\end{pmatrix}, & \ab^{(1)}&:=\begin{pmatrix}
-\underline{x}^{(1)} \overline{x}^{(1)} \\
\vdots \\
-\underline{x}^{(s)} \overline{x}^{(s)}
\end{pmatrix}, \\
\Wb^{(2)}&:=\begin{pmatrix} 
-\Sb^{(1)} & \dots & -\Sb^{(s)}  
\end{pmatrix}, & \ab^{(2)}&:=0
\end{align*}
and the mapping $\hb_v(\xb) := \begin{pmatrix}
\xb & \xb^2 \end{pmatrix}^\top$.
\end{thm}

\vspace{1mm}
\begin{proof}
We obviously have
$$
(\underline{x}^{(i)}+\overline{x}^{(i)}) \xb - \xb^2 -\underline{x}^{(i)} \overline{x}^{(i)}=  (\xb-\underline{x}^{(i)})(\overline{x}^{(i)}-\xb).
$$
Hence, the proposed parametrization yields
\begin{equation}\label{eq:1D_quad_ReLU}
\!\,\,\Phib(\hb_v(\xb))=\!\sum\limits_{i=1}^s\!-\Sb^{(i)} \max \left\{0, (\xb-\underline{x}^{(i)}) (\overline{x}^{(i)}-\xb) \right\}\!.\!\!\!\!\!
\end{equation}
Clearly, $(\xb-\underline{x}^{(i)}) (\overline{x}^{(i)}-\xb)> 0$ if and only if the conditions
\begin{equation}
\label{eq:xBounds1}
   \xb-\underline{x}^{(i)} > 0 \quad \text{and} \quad \overline{x}^{(i)}-\xb > 0  
\end{equation}
or
\begin{equation}
\label{eq:xBounds2}
\xb-\underline{x}^{(i)} < 0 \quad \text{and} \quad \overline{x}^{(i)}-\xb < 0
\end{equation}
hold.
Now, conditions~\eqref{eq:xBounds1} are  equivalent to ${\xb \in \interior\big(\Rc^{(i)}\big)}$ and conditions~\eqref{eq:xBounds2} are infeasible for proper intervals with $\underline{x}^{(i)}<\overline{x}^{(i)}$. 
Since we further have $\interior\big(\Rc^{(i)}\big)\cap \interior\big(\Rc^{(j)}\big)=\emptyset$ whenever $i\neq j$, we obtain
$$
\Phib(\hb_v(\xb)) = \left\{
\begin{array}{lc}
-\Sb^{(1)} (\xb-\underline{x}^{(i)}) (\overline{x}^{(i)}-\xb)  & \text{if} \ \xb\in\Rc^{(1)}, \\
\quad\vdots & \vdots  \\
-\Sb^{(s)} (\xb-\underline{x}^{(i)}) (\overline{x}^{(i)}-\xb) & \text{if} \ \xb\in\Rc^{(s)}, \\
\end{array}
\right.
$$
where $\Phib(\hb_v(\xb))=0$ results whenever $\xb$ is located on the boundary of any $\Rc^{(i)}$.  Evaluating \eqref{eq:deltaV} finally results in the PWA function
\begin{equation}
\label{eq:deltaV_PWA}
\Delta V_N(\xb) = \left\{
\begin{array}{lc}
\kappa^{(1)} \xb +\beta^{(1)}& \text{if} \ \xb\in\Rc^{(1)}, \\
\quad\vdots & \vdots  \\
\kappa^{(s)} \xb +\beta^{(s)}& \text{if} \ \xb\in\Rc^{(s)}\blind{,} \\
\end{array}
\right.    
\end{equation}
with the parameters $\kappa^{(i)}:=\lb^{(i)} + \Sb^{(i)} (\underline{x}^{(i)}+\overline{x}^{(i)})$ and $\beta^{(i)}:=c^{(i)} - \Sb^{(i)} \underline{x}^{(i)}\overline{x}^{(i)}$, which completes the proof.
\end{proof}

Roughly speaking, Theorem~\ref{thm:sol1} provides a ReLU-ANN that reflects the quadratic terms of $V_N(\xb)$. In fact, we proved that the difference~\eqref{eq:deltaV} is PWA. 
Now, as summarized in Section~\ref{subsection:PWA_ANN}, it has been known before that PWA functions can likewise be described by ReLU-ANN with one hidden layer. Thus, by combining both results, we can construct a ReLU-ANN of the form \eqref{eq:ReLUoneLayer} that completely reflects $V_N(\xb)$. However, when applying the results from Section~\ref{subsection:PWA_ANN}, one has to take into account that representing the quadratic terms requires $\xib=\hb_v(\xb)$. Fortunately, $\hb_v(\xb)$ contains the state~$\xb$. Hence, we can easily adapt the parametrization from~\eqref{eq:weightsBiasesPWA} according to the following Corollary.

\begin{cor}
\label{cor:ANNforDeltaV}
Let $\hb_v$ and $\Delta V_N$ be as in Theorem~\ref{thm:sol1} and \eqref{eq:deltaV_PWA}, respectively. Then,  $\Delta V_N(\xb)=\Phib(\hb_v(\xb))$ holds for an ANN as in \eqref{eq:ReLUoneLayer} with the parameters
\begin{align*}
 \Wb^{(1)}&:=\begin{pmatrix}
-1 & 0\\ 
\blind{+}1 & 0 \\ 
\blind{+}\vdots & \vdots \\ 
\blind{+}1 & 0 \\ 
\end{pmatrix},
\qquad 
\ab^{(1)}:=\begin{pmatrix}
\blind{+}\overline{x}^{(1)} \\ 
-\overline{x}^{(1)} \\ 
\vdots \\ 
-\overline{x}^{(s-1)} 
\end{pmatrix}, \\
\nonumber
\Wb^{(2)}&:=\begin{pmatrix}-\kappa^{(1)} & \!\kappa^{(2)} & \kappa^{(3)}\!-\!\kappa^{(2)} & \!\!\dots\!\!\! &  \kappa^{(s)}\!-\!\kappa^{(s-1)} \end{pmatrix}\!,\\
\nonumber
\ab^{(2)}&:=\kappa^{(1)}\overline{x}^{(1)} + \beta^{(1)}.
\end{align*}
\end{cor}

\vspace{2mm}
The combination of Theorem~\ref{thm:sol1} and Corollary \ref{cor:ANNforDeltaV} leads to the following summarizing statement. An illustration with an example will follow in Section \ref{subsec:ex_OVF}.

\begin{cor}
\label{cor:ANNforV}
Let $n=1$. Then, $V_N$ as in~\eqref{eq:VPWQ} can be exactly described by a ReLU-ANN with one hidden layer of width $w_1=2s$ and the inputs
 $\xib=\begin{pmatrix} \xb & \xb^2 \end{pmatrix}^\top$\!.
\end{cor}

\subsection{Conditional solution to Problem 2 for arbitrary $n$}

Interestingly, the Problems~\ref{prob:ANNforOVF} and~\ref{prob:ANNforQfun} are closely related. In fact, the following theorem shows that a solution to the former problem implies a solution to the latter. In other words, an ANN representing the OVF can be easily adapted for describing the Q-function.  

\begin{thm}
\label{thm:sol1ImpliesSol2}
Assume Problem~\ref{prob:ANNforOVF} has been solved for $N^\prime=N-1$ and let $\Wb^{(1)}_v$,  $\Wb^{(2)}_v$, $\ab_v^{(1)}$, and $\ab_v^{(2)}$ denote the corresponding weights and biases in~\eqref{eq:ReLUoneLayer}. Then, a solution to Problem~\ref{prob:ANNforQfun} is given in terms of 
\begin{subequations}
\label{eq:Wa_QFun}
\begin{align}
  \Wb^{(1)}&:=\begin{pmatrix}
\Wb^{(1)}_v & \zerob \\
\zerob & 1
\end{pmatrix}, & \ab^{(1)}&:=\begin{pmatrix}
\ab^{(1)}_v  \\
0
\end{pmatrix},\\
\Wb^{(2)}&:=\begin{pmatrix}
\Wb^{(2)}_v & 1
\end{pmatrix}, & \ab^{(2)} &:=\ab^{(2)}_v  
\end{align}
\end{subequations}
and the mapping
\begin{equation}
\label{eq:hqNaiv}
  \hb_q(\xb,\ub):=\begin{pmatrix}
\hb_v(\Ab\xb+\Bb \ub) \\
\ell(\xb,\ub)  
\end{pmatrix}.
\end{equation}
\end{thm}

\begin{proof}
The assumed solution to Problem~\ref{prob:ANNforOVF} implies 
\begin{align*}
 V_{N-1}(\xb)&=\Wb_v^{(2)} \max\left\{ \zerob, \Wb_v^{(1)} \hb_v(\xb) + \ab_v^{(1)} \right\}  \\
 &\quad + \ab_v^{(2)} +\Delta V_{N-1}(\xb) 
\end{align*}
with $\Delta V_{N-1}(\xb)$ being PWA in $\xb$. Hence, the Q-function \eqref{eq:Q-function} can be written as
\begin{align*}
Q_N(\xb,\ub) &= \Wb_v^{(2)} \max\left\{ \zerob, \Wb_v^{(1)} \hb_v(\Ab\xb+\Bb \ub) + \ab_v^{(1)} \right\} \\
&\quad + \ab_v^{(2)} +\Delta V_{N-1}(\Ab\xb+\Bb \ub) + \ell(\xb,\ub). 
\end{align*}
Now, the proposed solution to~\ref{prob:ANNforQfun} involves the ANN
$$
\Phib(\hb_q(\xb,\ub))=\Wb^{(2)} \max\left\{ \zerob, \Wb^{(1)} \hb_q(\xb,\ub) + \ab^{(1)} \right\} + \ab^{(2)}.
$$
Hence, substituting the proposed weights and biases and evaluating \eqref{eq:deltaQ} results in
\begin{align}
\nonumber
\Delta Q_N(\xb,\ub) & = \Delta V_{N-1}(\Ab\xb+\Bb \ub) + \ell(\xb,\ub)\\ 
\label{eq:deltaQViaDeltaV}
&\quad -\max\{0,\ell(\xb,\ub)\}. 
\end{align}
Clearly, $\Delta V_{N-1}(\Ab\xb+\Bb \ub)$ is PWA in $\xb$ and $\ub$. This completes the proof since $\ell(\xb,\ub) -\max\{0,\ell(\xb,\ub)\}=0$ due to positive semi-definiteness of $\ell$.
\end{proof}

From a theoretical point of view, Theorem~\ref{thm:sol1ImpliesSol2} is quite useful since it closely relates Problems~\ref{prob:ANNforOVF} and~\ref{prob:ANNforQfun}. Unfortunately, from a practical point of view, the proposed mapping~\eqref{eq:hqNaiv} is not convenient. In fact, $Q$-learning allows model-free learning but evaluating~\eqref{eq:hqNaiv} (for data preparation) requires a model in terms of $\Ab$ and~$\Bb$. Fortunately, this issue can be easily solved. 
To this end, we assume that the solution to Problem~\ref{prob:ANNforOVF} builds on the mapping 
\begin{equation}
\label{eq:hvStructure}
   \hb_v(\xb):=\begin{pmatrix}
\xb^\top & \xb_1^2 &\xb_1 \xb_2 & \!\dots\! & \xb_1 \xb_n  & \!\dots\!& \xb_n^2
\end{pmatrix}^\top\!\!\!,\!
\end{equation}
i.e., an extension of the state $\xb$ by all possible products of two individual states. We note that this assumption is reasonable for various reasons. First, \eqref{eq:hvStructure} includes our solution for the one-dimensional case in Theorem~\ref{thm:sol1}. Second, similar approaches have been considered for (loosely) related problems \cite{Bradtke1994,Fan2019}. Third, we show in Section~\ref{subsection:2D_example} that \eqref{eq:hvStructure} also applies to exemplary extensions with $n>1$. Now, the structure of~\eqref{eq:hvStructure} can be easily extended to the $\xb$-$\ub$-domain yielding the mapping
\begin{equation}
\label{eq:hqPrime}
  \hb^\prime_q(\xb,\ub):=\begin{pmatrix}
\hb_v(\xb) \\
\ub \\
\ub_1 \ub_2 \\
\vdots \\
\ub_m^2 \\
\xb_1 \ub_1 \\
\vdots \\
\xb_n \ub_m
\end{pmatrix}  
\end{equation}
that includes, in addtion to $\xb$ and $\ub$, all possible products of two entries in $(\,\xb^\top \,\,\,\ub^\top\,)$.
As shown next, there exists a linear relation between $\hb_q$ from Theorem~\ref{thm:sol1ImpliesSol2} and $\hb^\prime_q$ as in~\eqref{eq:hqPrime}.

\begin{lem}
\label{lem:hqReplacementViaL}
There exists a matrix $\Lb \in \R^{w_0 \times w_0^\prime}$ such that
\begin{equation}
\label{eq:hqViaL}
    \hb_q(\xb,\ub) = \Lb \hb^\prime_q(\xb,\ub),
\end{equation}
where $w_0:=n(n+3)/2+1$ and $w_0^\prime:=(n+m)(n+m+3)/2$.
\end{lem}

\begin{proof}
Clearly, each entry in ${\hb_v(\Ab \xb +\Bb \ub)}$ can be written as a linear combination of the entries in $\hb^\prime_q(\xb,\ub)$. The same observation holds for $\ell(\xb,\ub)$. The corresponding coefficients form the entries of $\Lb$. The dimensions $w_0$ and $w_0^\prime$ simply reflect the output dimensions of $\hb_q$ and $\hb^\prime_q$.
\end{proof}

Relation~\eqref{eq:hqViaL} allows to reformulate the statement in Theorem~\ref{thm:sol1ImpliesSol2} using the more convenient network inputs $\hb^\prime_q(\xb,\ub)$. In fact, as specified in the following corollary, the matrix $\Lb$ can simply be including in the construction~of~$\Wb^{(1)}$. 

\begin{cor}\label{cor:WL}
Consider the same solution to Problem~\ref{prob:ANNforOVF} as in Theorem~\ref{thm:sol1ImpliesSol2}. Further, let $\Lb$ satisfy~\eqref{eq:hqViaL}. 
Then, a solution to Problem~\ref{prob:ANNforQfun} can be obtained analogously to Theorem~\ref{thm:sol1ImpliesSol2} except that $\Wb^{(1)}$ and $\hb_q(\xb,\ub)$ are replaced by
$$
\Wb^{(1)}:=\begin{pmatrix}
\Wb^{(1)}_v & \zerob \\
\zerob & 1
\end{pmatrix} \Lb 
$$
and $\hb^\prime_q(\xb,\ub)$, respectively.
\end{cor}

Inspired by Corollaries \ref{cor:ANNforDeltaV} and \ref{cor:ANNforV}, we could now aim for statements about suitable ReLU-ANN for describing $\Delta Q_N$ and~$Q_N$, respectively.
However, Theorem~\ref{thm:sol1ImpliesSol2} assumes a solution to Problem~\ref{prob:ANNforOVF} and it is unclear whether such a solution exists for $n>1$ and, if it exists, how it is structured.
Thus, we only provide the following statement for $n=1$.

\begin{cor}
\label{cor:ANNforQ}
Let $n=1$ and assume $V_{N-1}$ is defined on~$s^\prime$ regions. Then, $Q_N$ as in~\eqref{eq:Q-function} can be exactly described by a ReLU-ANN with one hidden layer of width $w_1=2s^\prime+1$ and the inputs
 $\xib=\hb_q^\prime(\xb,\ub)$.
\end{cor}

\begin{proof}
We know from Theorem~\ref{thm:sol1} that Problem~\ref{prob:ANNforOVF} can be solved for $N^\prime=N-1$ based on a ReLU-ANN with one hidden layer of width $s^\prime$. Now, Theorem~\ref{thm:sol1ImpliesSol2} tells us that adding one neuron allows us to solve Problem~\ref{prob:ANNforQfun}. From~\eqref{eq:deltaQViaDeltaV}, we further infer that the resulting PWA $\Delta Q_N(\xb,\ub)$ is equivalent to $\Delta V_{N-1}(\Ab\xb+\Bb \ub)$. Corollary~\ref{cor:ANNforDeltaV} shows that another $s^\prime$ neurons are required to represent $\Delta V_{N-1}$, which already leads to the proposed width of $2s^\prime+1$. It remains to comment on suitable inputs for the ANN. In this context, we first note that $\hb_q(\xb,\ub)$ from Theorem~\ref{thm:sol1ImpliesSol2} contains $\Ab \xb + \Bb \ub$ for $\hb_v$ as in~\eqref{eq:hvStructure}. Hence, $\Delta V_{N-1}(\Ab\xb+\Bb \ub)$ can be described based on the ReLU-ANN for $\Delta V_{N-1}$ with the inputs $\hb_q(\xb,\ub)$. This completes the proof since Lemma~\ref{lem:hqReplacementViaL} implies that the inputs $\hb_q(\xb,\ub)$ can easily be replaced by $\xib=\hb_q^\prime(\xb,\ub)$ without changing the width $w_1$.
\end{proof}

\section{Numerical examples}
\label{sec:example}
We illustrate (and extend) our results with four numerical examples. The two first examples demonstrate the application of Theorems~\ref{thm:sol1} and~\ref{thm:sol1ImpliesSol2}. Moreover, we point out the benefits of the proposed ANN for learning the Q-function. The last example shows a possible direction for an open extension to systems with $n>1$.

\vspace{-1mm}

\subsection{Exact representation of the optimal value function}\label{subsec:ex_OVF}

\begin{figure}[h]
\centering
\includegraphics[trim={3cm 11cm 3cm 11cm},clip,scale=0.5]{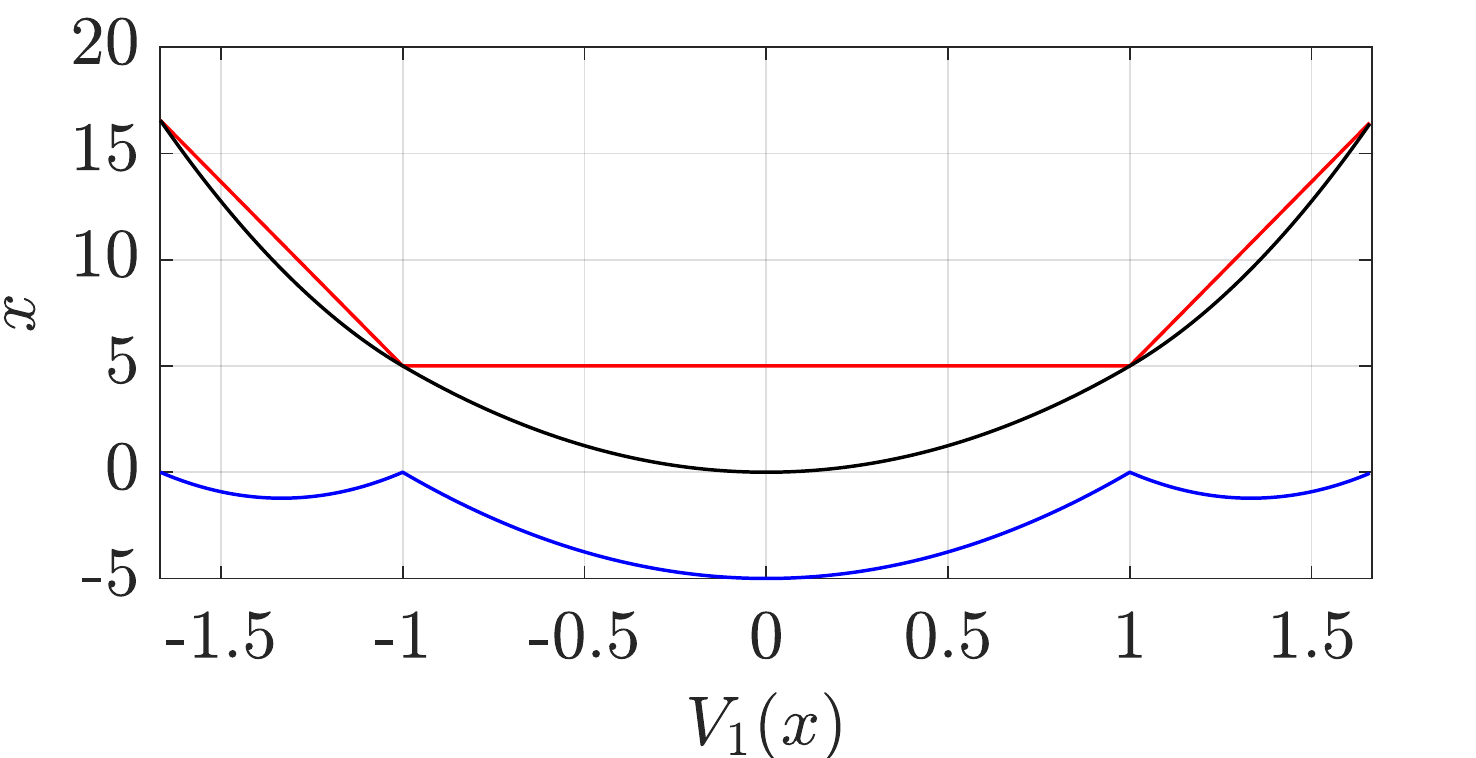}
\caption{Illustration of the PWQ function $V_1$ (black), the proposed ANN for the quadratic terms (blue), and the PWA difference $\Delta V_1$ (red) for the example in Section~\ref{subsec:ex_OVF}.}
\label{fig:V1}
\end{figure}

We consider the system from \cite[Ex.~2]{SchulzeDarup2016_ECC_MPC}
with the dynamics
\begin{equation}\label{eq:example_1D_system}
\xb(k+1)=\tfrac{6}{5}\xb(k)+\ub(k)\\
\end{equation}
and the constraints $\xb \in [-10,10]\subset \R$ and $\ub  \in [-1,1]\subset \R$.  As in \cite{SchulzeDarup2016_ECC_MPC}, we choose $\Qb=19/5$, $\Rb=1$, $\Pb=5$, and $\Tc=[-1,1]$. Finally, we select $N=1$ for illustration purposes here.
Explicitly solving \eqref{eq:OCP} then leads to
\begin{equation}\label{eq:example1_V1}
V_1(\xb) = \left\{
\begin{array}{lll}
11\xb^2+12\xb+6 & \text{if} & \xb\in\Rc^{(1)}, \\
5\xb^2 & \text{if} & \xb\in\Rc^{(2)}, \\
11\xb^2-12\xb+6 & \text{if} & \xb\in\Rc^{(3)}, 
\end{array}
\right. 
\end{equation}
where the three regions refer to the intervals 
$$
  \Rc^{(1)}=\left[-\tfrac{5}{3},-1\right], \quad \Rc^{(2)}=\left[-1,1\right], \quad
\Rc^{(3)}=\left[1,\tfrac{5}{3}\right].   
$$
Now, according to Theorem~\ref{thm:sol1}, a solution to Problem~\ref{prob:ANNforOVF} is given in terms of the parameters
\begin{subequations}
\label{eq:Wa_Ex1_V1}
\begin{align}
\Wb_v^{(1)}&:=\begin{pmatrix}
-\frac{8}{3} & -1 \\
\blind{+} 0 & -1 \\
\blind{+} \frac{8}{3} & -1 \\
\end{pmatrix}, &
\ab_v^{(1)}&:=\begin{pmatrix}
-\frac{5}{3} \\
\blind{+} 1 \\
-\frac{5}{3} \\
\end{pmatrix}, \\
\Wb_v^{(2)}&:=\begin{pmatrix}
-11 & -5 & -11 
\end{pmatrix}, &
\ab_v^{(2)} &:= 0
\end{align}
\end{subequations}
and the mapping $\hb_v(\xb):=(\,\xb\,\,\,\xb^2\,)^\top$. Figure~\ref{fig:V1} confirms this result by visualizing the PWA $\Delta V_1$.
We can further exploit the structure of $\Delta V_1$ based on Corollary~\ref{cor:ANNforDeltaV}. In fact, with $\kappa^{(i)}$ and $\beta^{(i)}$ as in the proof of Theorem~\ref{thm:sol1}, we find that $\Delta V_1$ can be described as an ANN of the form~\eqref{eq:ReLUoneLayer} with the parameters
\begin{subequations}
\label{eq:Wa_DeltaV_Ex1}
\begin{align}
\Wb_\Delta^{(1)}&:=\begin{pmatrix}
-1 & 0\\
\blind{+} 1 & 0 \\
\blind{+} 1 & 0
\end{pmatrix}, &
\ab_\Delta^{(1)}&:=\left(\begin{array}{c}
-1 \\
\blind{+} 1 \\
-1
\end{array}\right), \\
\Wb_\Delta^{(2)}&:=\left(\begin{array}{ccccc}
 \frac{52}{3} & 0 & \frac{52}{3}  
\end{array}\right), & \ab_\Delta^{(2)} &:= 5.
\end{align}
\end{subequations}
By combining both results, we can construct a ReLU-ANN that exactly describes $V_1$. In fact, the parametrization
\begin{align*}
\Wb^{(1)}&:=\begin{pmatrix}
\Wb_v^{(1)} \\
\Wb_\Delta^{(1)}
\end{pmatrix}, &
\ab^{(1)}&:=\begin{pmatrix}
\ab_v^{(1)} \\
\ab_\Delta^{(1)}
\end{pmatrix}, \\
\Wb^{(2)}&:=\begin{pmatrix}
\Wb_v^{(2)} & \Wb_\Delta^{(2)} 
\end{pmatrix}, &
\ab^{(2)} &:= \ab_\Delta^{(2)}
\end{align*}
leads to $V_1(\xb)=\Phib(\hb_v(\xb))$ and thus confirms Corollary~\ref{cor:ANNforV}.

\vspace{-5mm}

\subsection{Adaptation for Q-function}\label{subsec:ex_Q-function}

We next illustrate the application of Theorem~\ref{thm:sol1ImpliesSol2}. In order to reuse the results from the previous example, we consider $N=2$ here and consequently the Q-function
\begin{equation}
Q_2(\xb,\ub)=\ell(\xb,\ub) + V_1\left(\tfrac{6}{5} \xb + \ub\right)
\end{equation}
with $V_1$ as in~\eqref{eq:example1_V1}.
Now, according to Theorem~\ref{thm:sol1ImpliesSol2}, a solution to Problem~\ref{prob:ANNforQfun} can be obtained by substituting~\eqref{eq:Wa_Ex1_V1} in~\eqref{eq:Wa_QFun} and by considering the mapping
$$
\hb_q(\xb,\ub):=\begin{pmatrix}
\hb_v\big(\frac{6}{5} \xb + \ub\big) \\[1mm]
\ell(\xb,\ub)
\end{pmatrix} = \begin{pmatrix}
\frac{6}{5} \xb + \ub \\
\frac{36}{25} \xb^2 + \frac{12}{5} \xb \ub + \ub^2 \\
\frac{19}{5} \xb^2 + \ub^2 
\end{pmatrix}.
$$
Clearly, evaluating this mapping requires the model~\eqref{eq:example_1D_system}.
Fortunately, Lemma~\ref{lem:hqReplacementViaL} and Corollary~\ref{cor:WL} tell us that we can easily switch to the more convenient network inputs
\begin{equation}
\label{eq:hqPrime_Ex2}
\hb^\prime_q(\xb,\ub):=\begin{pmatrix}
\xb &
\xb^2 &
\ub &
\ub^2 &
\xb \ub 
\end{pmatrix}^\top\!. 
\end{equation}
To this end, we quickly verify that~\eqref{eq:hqViaL} is satisfied by
$$
\Lb =\begin{pmatrix}
\frac{6}{5} & 0 & 1 & 0 & 0 \\[.5ex]
0 & \frac{36}{25} & 0 & 1 & \frac{12}{5} \\[.5ex]
0 & \frac{19}{5} & 0 & 1 & 0 \\
\end{pmatrix}.
$$
Hence, Theorem~\ref{thm:sol1ImpliesSol2} in combination with the parameters~\eqref{eq:Wa_Ex1_V1} and Corollary~\ref{cor:WL} implies that Problem~\ref{prob:ANNforQfun} is solved by
\begin{align*}
\Wb_q^{(1)}\!&:=\!\begin{pmatrix}
-\frac{16}{5} & \!-\frac{36}{25} & \!-\frac{8}{3} & \!-1 & -\frac{12}{5} \\[.5ex]
\blind{+}0 & \!-\frac{36}{25} & \!\blind{+} 0 & \!-1 & -\frac{12}{5} \\[.5ex]
\blind{+}\frac{16}{5} & \!-\frac{36}{25} & \!\blind{+}\frac{8}{3} & \!-1 & -\frac{12}{5} \\[.5ex]
\blind{+}0 & \!\blind{+}\frac{19}{5} & \!\blind{+}0 & \!\blind{+}1 & \blind{+} 0
\end{pmatrix}\!, &
\!\!\!\ab_q^{(1)}&:=\begin{pmatrix}
-\frac{5}{3} \\
\blind{+} 1 \\
-\frac{5}{3} \\
\blind{+} 0 \\
\end{pmatrix}\!, \\
\Wb_q^{(2)}\!&:=\begin{pmatrix}
-11 & -5 & -11 & 1
\end{pmatrix}, &
\!\!\!\ab_q^{(2)} &:= 0.
\end{align*}
and the mapping~\eqref{eq:hqPrime_Ex2}.
In order to describe the entire Q-function, it remains to construct an ANN for $\Delta Q_2(\xb,\ub)=\Delta V_1(\Ab \xb+ \Bb \ub)$. Clearly, we can reuse the parametrization for $\Delta V_1$ from~\eqref{eq:Wa_DeltaV_Ex1}. However, we have to include a matrix that selects $\Ab \xb+ \Bb \ub$ from $\hb_q$ and we have to take into account the switching to $\hb_q^\prime$ via $\Lb$. Doing so, we can easily confirm that $\Delta V_1(\Ab \xb+ \Bb \ub)=\Phib(\hb_q^\prime(\xb,\ub))$ holds for
$$
\Wb^{(1)}_\delta:=
\Wb^{(1)}_\Delta \begin{pmatrix} 1 & 0 & 0 \\
1 & 0 & 0 \\
\end{pmatrix} \Lb=
\begin{pmatrix}
-\frac{6}{5} & 0 & -1 & 0 & 0 \\[.5ex]
\blind{+}\frac{6}{5} & 0 & \blind{+}1 & 0 & 0 \\[.5ex]
\blind{+}\frac{6}{5} & 0 & \blind{+}1 & 0 & 0 \\
\end{pmatrix},
$$
$$\ab^{(1)}_\delta:=\ab^{(1)}_\Delta, \ \Wb^{(2)}_\delta:=\Wb_\Delta^{(2)}, \ \text{and} \ \ab^{(2)}_\delta:=\ab^{(2)}_\Delta.$$
Combining the solution to Problem~\ref{prob:ANNforQfun} and the description of $\Delta Q_2$ from above leads to a ReLU-ANN for $Q_2$ with the parameters 
\begin{align*}
\Wb^{(1)}&:=\begin{pmatrix}
\Wb_q^{(1)} \\
\Wb_\delta^{(1)}
\end{pmatrix}, &
\ab^{(1)}&:=\begin{pmatrix}
\ab_q^{(1)} \\
\ab_\delta^{(1)}
\end{pmatrix}, \\
\Wb^{(2)}&:=\begin{pmatrix}
\Wb_q^{(2)} & \Wb_\delta^{(2)} 
\end{pmatrix}, &
\ab^{(2)} &:= \ab_\delta^{(2)}
\end{align*}
and the inputs $\hb_q^\prime(\xb,\ub)$. As predicted by Corollary~\ref{cor:ANNforQ}, this ANN has one hidden layer of width $w_1=2s^\prime+1=7$.

\subsection{Learning of the Q-function}\label{subsection:example_learn_Q}

The presented work is motivated by the observation that, in the context of learning, ANN topologies are often chosen based on trial-and-error. Our results show that, for MPC-related learning, more educated choices exist that even allow to exactly describe the functions of interest. Interestingly, the identified topologies (summarized in the Corollaries~\ref{cor:ANNforV} and~\ref{cor:ANNforQ}) seem to be useful beyond the proposed exact descriptions. To substantiate this observation, we investigated different learning-based approximations of $Q_2$ from the previous example. More precisely, we trained various ReLU-ANN with different inputs, widths, and depths (using supervised learning and Matlab's Deep Learning Toolbox \cite{MatlabDeepLearning} for simplicity) over $1000$ epochs. We repeated the training 100 times for each ReLU-ANN, where we used the same randomly sampled points from the Q-function as training data for every trial. At the end of every trial, we computed the root-mean-square-error (RMSE). The mean values of the resulting RMSE are listed in Table~\ref{tab:RMSE}. As apparent from the data, the proposed topology with one hidden layer ($l=1$) of width $w_1=7$ and the inputs $\hb_q^\prime(\xb,\ub)$ performed best even though some ANN incorporated more parameters. 

\begin{table}[h]
\caption{RMSE for different ReLU-ANN (where $w_i$ refers to $w_1,\dots,w_l$ and $\#_p$ reflects the number of parameters).}
\centering
\begin{tabular}{crrrr}
\toprule
\quad$\xib$\quad\quad &\qquad$l$ &\qquad$w_i$ &\qquad$\#_p$ &\qquad RMSE \\ 
\midrule
$\hb^\prime_q(\xb,\ub)$ & $1$ & $5$ & $36$ & $0.4084$ \\[.5ex]
$\hb^\prime_q(\xb,\ub)$ & $1$ & $6$ & $43$ & $0.2589$ \\[.5ex] 
$\hb^\prime_q(\xb,\ub)$ & $1$ & $7$ & $50$ & $0.1854$ \\[.5ex] 
$\hb^\prime_q(\xb,\ub)$ & $1$ & $8$ & $57$ & $0.1999$ \\[.5ex] 
$(\,\xb\,\,\,\ub\,)^\top$ & $1$ & $12$ & $49$ & $0.2065$ \\[.5ex]
$(\,\xb\,\,\,\ub\,)^\top$ & $2$ & $5$ & $51$ & $0.3092$ \\[.5ex] 
$(\,\xb\,\,\,\ub\,)^\top$ & $3$ & $4$ & $57$ &  $0.2735$\\
\bottomrule
\label{tab:RMSE}
\end{tabular}
\end{table}

\vspace{-5mm}

\subsection{Extension to two-dimensional case}\label{subsection:2D_example}

\begin{figure}[tp]
\centering
\includegraphics[trim={3cm 9cm 3cm 10cm},clip,scale=0.5]{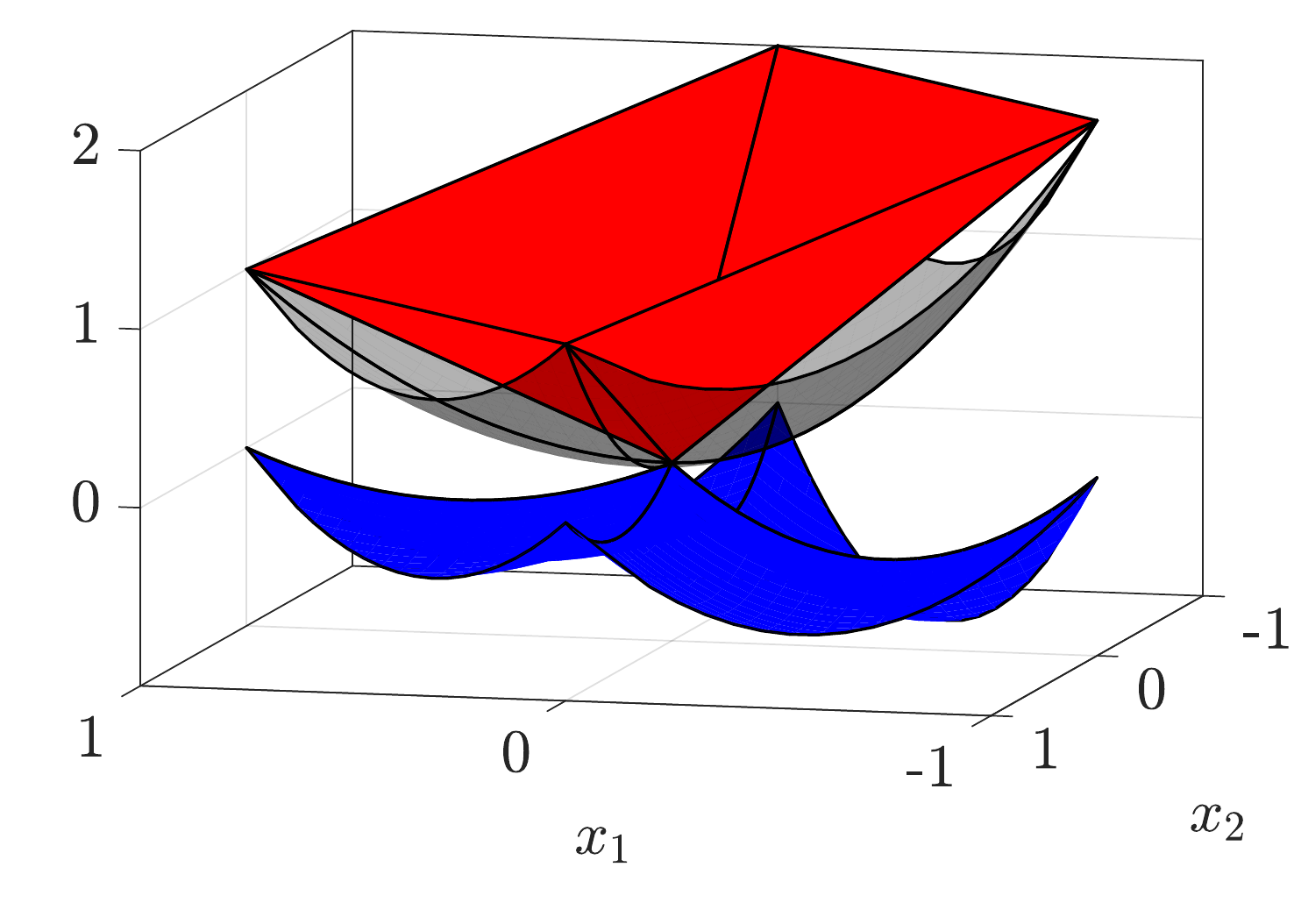}
\caption{Illustration of the PWQ function $V_N$ (grey), the proposed ANN for the quadratic terms (blue), and the PWA difference $\Delta V_N$ (red) for the example in Section~\ref{subsection:2D_example}.}
\label{fig:PWQ_2D}
\end{figure}

We next show that solutions to Problem~\ref{prob:ANNforOVF} may also exist for $n>1$ and that they may be derived using strategies similar to those underlying our solution for $n=1$.
To this end, we consider the fictive function
\begin{equation}\label{eq:example_2D_PWQ}
V_N(\xb) := \left\{
\begin{array}{ll}
\blind{2}\xb_1^2+\xb_2^2 & \text{if}\,\,\,\xb\in\Rc^{(1)}, \\
2\xb_1^2+\xb_2^2 & \text{if}\,\,\,\xb\in\Rc^{(2)}, \\
2\xb_1^2+2\xb_2^2 & \text{if}\,\,\,\xb\in\Rc^{(3)}, \\
\blind{2}\xb_1^2+2\xb_2^2 & \text{if}\,\,\,\xb\in\Rc^{(4)} \\
\end{array}
\right. 
\end{equation}
(without specifying $N$) defined on the regions
\begin{align*}
    \Rc^{(1)}&:=\{\xb \in \R^2 \,|\, \xb_1 \geq 0,\, \xb_2 \geq 0,\, \blind{+}\xb_1+\xb_2 \leq 1\}, \\
    \Rc^{(2)}&:=\{\xb \in \R^2 \,|\, \xb_1 \leq 0,\, \xb_2 \geq 0,\, -\xb_1+\xb_2 \leq 1\}, \\
    \Rc^{(3)}&:=\{\xb \in \R^2 \,|\, \xb_1 \leq 0,\, \xb_2 \leq 0,\, -\xb_1-\xb_2 \leq 1\}, \\
    \Rc^{(4)}&:=\{\xb \in \R^2 \,|\, \xb_1 \geq 0,\, \xb_2 \leq 0,\, \blind{+}\xb_1-\xb_2 \leq 1\}. 
\end{align*}
We next show that 
\begin{align}
\nonumber
\Phib(\hb_v(\xb))=&- \max\{0 , \xb_1 (-\xb_1 +\xb_2-1) \} \\
\nonumber
&- \max\{0 , \xb_2 (-\xb_1 -\xb_2-1) \} \\
\nonumber
&- \max\{0 , \xb_1 (-\xb_1 -\xb_2-1) \} \\
\nonumber
&- \max\{0 , \xb_2 (\xb_1 -\xb_2-1) \} \\
\nonumber
&-0.5 \max\{0 , -\xb_2 (\xb_1 +\xb_2-1) \} \\
\nonumber
&-0.5 \max\{0 , -\xb_1 (\xb_1 +\xb_2-1) \} \\
\nonumber
&-0.5 \max\{0 , -\xb_2 (-\xb_1 +\xb_2-1) \} \\
\label{eq:Phi_example_2D}
&-0.5 \max\{0 , -\xb_1 (\xb_1 -\xb_2-1) \}
\end{align}
with $\hb_v$ as in~\eqref{eq:hvStructure} represents a solution to Problem~\ref{prob:ANNforOVF}. In this context, we first note that $V_N(\xb)-\Phib(\hb_v(\xb))$ is indeed PWA as visualized in Figure~\ref{fig:PWQ_2D}.
Furthermore, $\Phib$ equals~\eqref{eq:ReLUoneLayer} with the parameters
\begin{equation}
\nonumber
\Wb^{(1)}\!=\begin{pmatrix}
-1 & \blind{+}0 & -1 & \blind{+}1 & \blind{+}0 \\
\blind{+}0 & -1 & \blind{+}0 & -1 & -1 \\
-1 & \blind{+}0 & -1 & -1 & \blind{+}0 \\
\blind{+}0 & -1 & \blind{+}0 & \blind{+}1 & -1 \\
\blind{+}0 & \blind{+}1 & \blind{+}0 & -1 & -1 \\
\blind{+}1 & \blind{+}0 & -1 & -1 & \blind{+}0 \\
\blind{+}0 & \blind{+}1 & \blind{+}0 & \blind{+}1 & -1 \\
\blind{+}1 & \blind{+}0 & -1 & \blind{+}1 & \blind{+}0 \\
\end{pmatrix},\,\, 
\Wb^{(2)}\!=\begin{pmatrix}
-1 \\
-1 \\
-1 \\
-1 \\
-0.5 \\
-0.5 \\
-0.5 \\
-0.5 \\
\end{pmatrix}^{\!\top}\!\!\!,
\end{equation}
$\ab^{(1)}=\zerob$, and $\ab^{(2)}=0$ and the inputs $\xib:=\hb_v(\xb)$.
Hence, \eqref{eq:Phi_example_2D} is indeed a solution to Problem~\ref{prob:ANNforOVF}.

We briefly comment on the derivation of this solution. First, it is easy to see that every ReLU in~\eqref{eq:Phi_example_2D} is constructed in such a way that it is either active or inactive on the individual regions $\Rc^{(i)}$. For instance, $\max\{0 , \xb_1 (-\xb_1 +\xb_2-1) \}$ is active on $\Rc^{(2)},\Rc^{(3)}$ and inactive on $\Rc^{(1)},\Rc^{(4)}$. In total, four ReLU are active on every region. Since we can freely choose the eight weighting factors of the ReLU, we can use these degrees of freedom to eliminate the eight quadratic terms in~\eqref{eq:example_2D_PWQ}. The strategy is closely related to the proof of Theorem~\ref{thm:sol1} and can, in principle, be generalized. However, formulating and solving the underlying system of equations is highly non-trivial.

\section{Conclusions and Outlook}
\label{sec:conclusion}

In this paper, we presented methods for exactly describing PWQ functions using tailored ANN with ReLU activations. In particular, we showed that the OVF in linear MPC can always be described by a ReLU-ANN with one hidden layer for the special case $n=1$ (see Thm.~\ref{thm:sol1} and Cor.~\ref{cor:ANNforV}). Moreover, according Theorem~\ref{thm:sol1ImpliesSol2}, a given description for the OVF can always be modified for exactly representing the associated Q-function (for any $n\geq1$).

We illustrated our results with various numerical examples. Two observations are of particular interest. First, the example in Section~\ref{subsection:example_learn_Q} indicates an advantage of the proposed ANN topologies for learning the Q-function. Second, the example in Section~\ref{subsection:2D_example} shows that the approach for representing the OVF might be extendable to $n>1$. Both observations offer interesting directions for further research.

% References

\end{document}